\documentclass[aps,prb,superscriptaddress,twocolumn,showpacs]{revtex4}

\usepackage{amsmath,amssymb}
\usepackage{graphicx}
\usepackage{feynmf}
\usepackage{array}	

\newcommand{\rs}{\rm \scriptscriptstyle}
\newcommand{\s}{\scriptscriptstyle}

\begin{document}

\title{Scattering matrix approach to interacting electron transport}

\author{D. Oehri}
\affiliation{Theoretische Physik, Wolfgang-Pauli-Strasse 27,
  ETH Zurich, CH-8093 Z\"urich, Switzerland}

\author{A.V. Lebedev}
\affiliation{Theoretische Physik, Wolfgang-Pauli-Strasse 27,
  ETH Zurich, CH-8093 Z\"urich, Switzerland}

\author{G.B. Lesovik}
\affiliation{L.D.\ Landau Institute for Theoretical Physics RAS,
   117940 Moscow, Russia}

\author{G. Blatter}
\affiliation{Theoretische Physik, Wolfgang-Pauli-Strasse 27,
  ETH Zurich, CH-8093 Z\"urich, Switzerland}

\date{\today}

\begin{abstract}
We investigate the modification in mesoscopic electronic transport due to
electron-electron interactions making use of scattering states.  We
demonstrate that for a specific (finite range) interaction kernel, the
knowledge of the scattering matrix is sufficient to take interaction effects
into account. We calculate perturbatively the corrections to the current and
current-current correlator; in agreement with previous work, we find that, in
linear response, interaction effects can be accounted for by an effective
(renormalized) transmission probability.  Beyond linear response, simple
renormalization of scattering coefficients is not sufficient to describe the
current-current correlator, as additional corrections arise due to irreducible
two-particle processes.  Furthermore, we find that the correlations between
opposite-spin currents induced by interaction are enhanced for an asymmetric
scatterer, generating a nonzero result already to lowest order in the
interaction.
\end{abstract}


\maketitle

\section{Introduction}

Transport through mesoscopic systems in the presence of particle scattering
and interaction has been at the focus of experimental and theoretical studies
during several decades. The scattering matrix theory reducing the problem to
asymptotic states \cite{landauer:57,landauer:70,buettiker:86} proved to be an
efficient tool characterizing transport in the presence of single-particle
scattering and for non-interacting systems. Beyond calculation of the current,
the scattering matrix formalism was used \cite{lesovik:89,buettiker:90} to
determine the noise properties of devices and provided the complete
probability distribution of transmitted charge through the calculation of the
full counting statistics (FCS) \cite{levitov:93,levitov:96}.  Several reviews
dealing with the scattering matrix approach to electron transport are
available today \cite{blanter:00,lesovik:11}.

Effects of electron-electron interactions give rise to numerous interesting
phenomena, e.g., Coulomb blockade \cite{glazman:05}, Kondo effect
\cite{glazman:05}, or the 0.7 anomaly \cite{thomas:96} in transport, or
entanglement and decoherence within the context of quantum processing in
mesoscopic systems \cite{lerner:04}.  Interactions arise locally and thus
involve the particle wave function within the domain where interactions take
place.  A natural way to tackle the combined problem of scattering and
interactions is to start with the interacting problem (e.g., the quantum dot)
and include scattering thereafter, as in the tunneling approach which treats
the scattering part perturbatively.  In the present paper, we take an
alternative approach which describes the scattering part exactly for any
scatterer. We then have to compromise with the interaction, for which we
choose a particular kernel which is constant over the scattering region. This
scheme allows us to express the full interacting Hamiltonian through the exact
scattering coefficients of the non-interacting system. The approach can be
extended to the case of electron-photon/plasmon and electron-phonon
interactions, where the wave function of the scattered low-energy mode (rather
than the two-particle interaction kernel) is approximately constant over the
region of scattering and interaction. This idea has been exploited before in
the study of photon/plasmon generation by electron scattering, as discussed in
Refs.\ \onlinecite{beenakker:01,beenakker:04,lebedev:10}.

Accounting for interaction effects in electronic transport is a highly
non-trivial task and general results are sparse, with notable exceptions such
as the generalized Landauer formula due to Meir and Wingreen \cite{meir:92},
however.  Still, considerable progress can be made for specific systems:
Studies of the (Anderson \cite{anderson:61}) impurity model for quantum dots
provide non-perturbative results for the non-linear $I$-$V$ characteristics
\cite{saleur:02}, for the shot noise \cite{schiller:98}, and perturbative
results (and beyond) are available for the full counting statistics
\cite{gogolin:06}. The same quantities, current-voltage characteristic
\cite{mehta:06,boulat:08}, shot-noise \cite{branschaedel:10}, and full
counting statistics \cite{carr:11}, have been found for the interacting
resonant level model \cite{vigman:78,schlottmann:78}. The more general case of
a quantum point contact with an arbitrary two-particle interaction has been
analyzed with the help of field-theoretic methods: the current and noise have
been determined by Golubev {\it et al.} \cite{golubev:05} within a
saddle-point approximation and the renormalization group flow of the
transmission coefficients has been found by Kindermann and Nazarov
\cite{kindermann:03} in their study of weak interaction effects on full
counting statistics.  In our approach, we express the interaction through the
non-interacting scattering coefficients of an arbitrary mesoscopic conductor,
yet restricted to a capacitive type of Coulomb interaction.  Providing this
starting point, this interacting Hamiltonian can be used in a simple
perturbative calculation as done below, but could also be fed to a more
sophisticated technique as those developed in Refs.\
\onlinecite{golubev:05,kindermann:03}.

Much interest has been devoted to the question whether the Landauer formula
for the current survives in the presence of interactions, a question that our
approach is quite suitable to address.  For systems where the interaction
among electrons is restricted to a finite region of space, Meir and Wingreen
\cite{meir:92} developed a generalization of the Landauer formula for the
current, which was recently shown to reduce to the form of the Landauer
formula via introduction of an effective transmission \cite{ness:10}.
Furthermore, the analysis has been extended to include effects of interactions
in the leads \cite{ness:11}. Gogolin and Komnik \cite{gogolin:06}, devising a
method providing the generating function of the full counting statistics, have
found that, within linear response, effects of interaction in the Anderson
model can be accounted for with a renormalization of scattering coefficients.
However, it is still under debate whether all effects of interaction can be
included via an effective transmission probability in the Landauer formula
\cite{vignale:09}.

Below, we consider two (semi-infinite) non-interacting leads connected via a
finite region of space where single-electron scattering and interaction
between electrons takes place. We focus on an interaction between electrons
which is restricted to a finite region in space and furthermore constant in
the region of single-electron scattering. In this case, the matrix elements in
the interaction Hamiltonian can be calculated from the knowledge of the
asymptotic behavior of the single-particle wave-functions (scattering matrix).
Our study is based on the Keldysh formalism \cite{keldysh:65} and includes the
non-equilibrium situation with a finite applied bias voltage $V$. We treat the
scattering problem exactly by writing the initial density matrix
$\hat{\rho}(t_{0})$ (with $t_0\rightarrow -\infty$) in terms of the scattering
states of the non-interacting system.
Assuming an arbitrary scatterer, we limit ourselves to a perturbative
expansion in the interaction and stay with the lowest order (in which
case the introduction of Green's function as done below is not a must).
Adopting a specific scatterer, e.g., a single resonance level, we push the
perturbative expansion to higher order with a selected choice of diagrams.

In a first application of our approach, we determine the interaction-induced
correction to transport and noise: to lowest order in the interaction strength
$U_0$ and to all orders in the applied bias $V$, the corrections to the
single-particle Green's function $G$ and the transport current $j$ can be
described through renormalized scattering amplitudes. On the other hand, the
corrections to the two-particle Green's function and the current-noise cannot
be cast into the form of renormalized scattering coefficients in general.
This is due to the presence of irreducible two-particle processes (vertex
corrections).  Calculating the current and noise from the single- and
two-particle Green's function, we find that in the linear voltage regime $V$
all the effect of interaction can be accounted for by an effective
transmission (to order $U_0$), in agreement with previous findings by Gogolin
and Komnik\cite{gogolin:06}.  In the non-linear voltage regime, this is not
the case any more and interaction effects beyond a simple renormalization of
the scattering properties appear.  Furthermore, it turns out, that the corrections to the
two-particle Green's function generate a correlation between opposite-spin
currents in an asymmetric device which appears to order $U_0 V^2$, while in a
symmetric system\cite{gogolin:06} such a correlation arises only in order $U_0^2
V^2$. Again, our approach turns out to be quite suitable in unravelling such
additional correlations due to an asymmetry of the scatterer, a question that
has not been addressed before.

The paper is organized as follows: In Sec.~\ref{sec:gt}, we introduce the
Hamiltonian, briefly explain the non-equilibrium formalism used in the paper,
and demonstrate how the interaction matrix elements simplify due to our
particular choice of the interaction kernel. In Sec.~\ref{sec:c}, we calculate
the lowest order (in interaction) correction to the current and noise
correlator and find the interaction-induced correlation between opposite-spin
currents. We then evaluate the results for the specific case of a single
resonance level in Sec.~\ref{sec:srl}, including a resummation of diagrams
providing the mean-field correction to the single-particle Green's function.
A summary and conclusions are given in Sec.\ \ref{sec:sc}.

\section{Formalism}\label{sec:gt}

Our one-dimensional system involves two half-infinite leads connected
by a central region where particles are scattered and is described by
the Hamiltonian
\begin{align}
\hat{H}_{\s 0}=\hat{H}_{\rm kin}+\hat{V},
\end{align}
where the potential $V(x)$ gives rise to single-particle scattering. Making
use of the Lippmann-Schwinger (LS) scattering states $|\varphi_{\alpha
k}\rangle$, we treat the scattering problem exactly; the index $\alpha={\rm
L}/{\rm R}$ distinguishes left/right-incoming states and we choose
the wavevector $k>0$. The LS scattering states are eigenstates of the
Hamiltonian $\hat{H}_{\rs 0}$,
\begin{align}
\bigl(\hat{H}_{\rm kin}+\hat{V} \bigr)
\bigl| \varphi_{\alpha k}\bigr\rangle =
E_{k} \bigl| \varphi_{\alpha k}\bigr\rangle,
\end{align}
with energy $E_{k}$. In the following, we will consider a spectrum linearized
around $E_{\rs F}=\hbar \omega_{\rm\scriptscriptstyle F}$, i.e.,
$E_{k}=E_{\rm\scriptscriptstyle F}+\hbar v_{\rm\scriptscriptstyle F}
(k-k_{\rm\scriptscriptstyle F})$ (this is a convenient simplification rather
than a necessity).  The asymptotics of the LS states $(\varphi_{\alpha
k}(x)=\langle x | \varphi_{\alpha k} \rangle$),
\begin{equation}
\begin{aligned}
\varphi_{L k}(x)&=
\begin{cases}
e^{i k x}
+ r_{ L  k}\, e^{i k d} e^{-i k x},     & x\rightarrow -\infty, \\
t_{k}\,e^{i k d} e^{i k x},     	& x\rightarrow \phantom{-}\infty,
\end{cases}\\
\varphi_{R k}(x)&=
\begin{cases}
t_{k}\,e^{i k d} e^{-i k x},     & x\rightarrow -\infty, \\
e^{-i k x} +
r_{ R k}\,e^{i k d} e^{i k x},      & x\rightarrow \phantom{-}\infty,
\end{cases}
\end{aligned}
\label{eq:asymp}
\end{equation}
is determined by the scattering matrix
\begin{align}
S_{k}
=
\begin{pmatrix}
r_{Lk} & t_{k} \\
t_{k} & r_{Rk}
\end{pmatrix},
\label{eq:sm}
\end{align}
where $t_{Lk}=t_{Rk} = t_k$ in our time-reversal symmetric system. Different
from more standard formulations, we have defined the scattering coefficients
with additional phase factors $\exp(ikd)$ as this will simplify our
expressions later. The Lippmann-Schwinger states satisfy the orthogonality
\begin{align}
\langle \varphi_{\alpha k}|\varphi_{\beta q}\rangle
= 2\pi \delta(k-q)\delta_{\alpha\beta}
\label{eq:orth}
\end{align}
and, in the absence of bound states as assumed here, the condition of
completeness. The Hamiltonian $\hat{H}_{\s 0}$ then can be rewritten in terms
of the LS states as
\begin{align}
\hat{H}_{\s 0}=\sum_{\alpha,\sigma} \!
        \int \!\! \frac{dk}{2\pi}\, E_{k} \,
        \hat{c}^{\dagger}_{\alpha k \sigma}
        \hat{c}^{\phantom{\dagger}}_{\alpha k \sigma}
\label{eq:H0LS}
\end{align}
with the creation (annihilation) operator $\hat{c}^{\dagger}_{\alpha k
\sigma}$ ($\hat{c}^{\phantom{\dagger}}_{\alpha k \sigma}$) of LS scattering
states with spin $\sigma$.  With a voltage $V$ applied to the system, we
assume a non-equilibrium steady state at an initial time $t_{\rs 0}$ ($t_{\rs
0}\rightarrow -\infty$) which we describe by the initial density matrix
\begin{align}
\hat{\rho}_{\rs 0}=\hat{\rho}(t_{\rs 0})=
\exp\Bigl[-\beta
\sum_{\alpha,\sigma}\int\! \frac{dk}{2\pi}
\bigl(E_{k}-\mu_{\alpha}\bigr)
c^\dagger_{\alpha k \sigma}
c^{\phantom{\dagger}}_{\alpha k \sigma}
\Bigr]
\label{eq:rho0}
\end{align}
with the chemical potential $\mu_{ L/R}=E_{\rm\scriptscriptstyle F}\pm eV/2$
in the left/right lead.

\subsection{Interaction matrix elements}
\label{sec:ime}

We consider an interaction between electrons described by the Hamiltonian
\begin{align}\label{eq:Hint_s}
\hat{H}_{\rm int}&=\frac12\sum_{\sigma\sigma^\prime}\int\! dx\,
dx^\prime\,
        \hat{\psi}_{\sigma}^\dagger(x)
        \hat{\psi}_{\sigma^\prime}^\dagger(x^\prime) \\ \nonumber
        & \qquad\qquad\qquad \times U(x,x^\prime)
        \hat{\psi}_{\sigma'}^{\phantom{\dagger}}(x^\prime)
        \hat{\psi}_{\sigma}^{\phantom{\dagger}}(x)
\end{align}
with $U(x,x')$ the interaction kernel. Here, we focus on a situation where the
interaction can be assumed constant within the region of single-particle
scattering. Typical physical systems we have in mind are quantum dots (with
constant interaction inside the dot and vanishing outside) and quantum point
contacts (where the interaction can be assumed large and constant between
electrons in the quantum point contact region).  Inspired by the capacitive
form $Q^2/2C$ (with the charge $Q$ and capacitance $C$) we choose an
electron-electron interaction kernel 
\begin{align}
U(x,x')=U_{\rs 0} f(x) f(x').
\label{eq:kernel}
\end{align}
Using the Lippmann-Schwinger scattering states, we express the field operators as
\begin{align}
\hat{\psi}_{\sigma}(x)=\sum_{\alpha} \! \int \frac{dk}{2\pi}\,
\varphi_{\alpha k}(x)\hat{c}_{\alpha k \sigma}
\end{align}
and can rewrite the Hamiltonian as
\begin{align}
\label{eq:Hint}
\hat{H}_{\rm int}&=\frac{U_{\rs 0}}{2}\sum_{\sigma\sigma^\prime}
\sum_{\alpha\alpha^\prime\beta\beta^\prime} \int \frac{dk}{2\pi}
\frac{dk^\prime}{2\pi}
\frac{dq}{2\pi} \frac{dq^\prime}{2\pi} \\
&\hspace{15pt} \times A_{\alpha^\prime k^\prime,\alpha k}
A_{\beta' q',\beta q} \, \hat{c}_{\alpha' k' \sigma}^\dagger
\hat{c}_{\beta' q' \sigma'}^\dagger \hat{c}_{\beta
q \sigma'}^{\phantom{\dagger}} \hat{c}_{\alpha k \sigma}^{\phantom{\dagger}}
\nonumber
\end{align}
with the matrix elements
\begin{align}
A_{\alpha k,\beta q}=\int dx f(x)
    \varphi_{\alpha k}^\ast(x)
    \varphi_{\beta q}^{\phantom{\ast}}(x).
\label{eq:ov1}
\end{align}
Assuming a scattering potential $V(x)$ which vanishes outside the region
$[-d/2,d/2]$, we consider a specific kernel with
\begin{align}
      f(x)=\Theta(d/2-|x|),
      \label{eq:f}
\end{align}
allowing us to rewrite the matrix elements in Eq.~(\ref{eq:ov1}) as
\begin{align}
      A_{\alpha k,\beta q}&= \langle \varphi_{\alpha k}|\varphi_{\beta q}\rangle
      -\int_{C_d} dx \,\,  \varphi_{\alpha k}^\ast(x)
      \varphi_{\beta q}^{\phantom{\ast}}(x)
\label{eq:ov2}
\end{align}
with $C_d=\mathbb{R}\,\backslash\, [-d/2,d/2]$.  The first term is given by
the orthogonality condition Eq.~(\ref{eq:orth}) of the LS states, while the
second term can be calculated using the asymptotic form Eq.~(\ref{eq:asymp})
of the LS states. As a result, the overlap matrix elements can be expressed
through the scattering coefficients alone, 
\begin{equation}
\begin{aligned}
A_{\alpha k,\alpha q}&=-i
\frac{1-
(r_{\alpha k}^\ast r_{\alpha q}^{\phantom{\ast}}
+t_{\alpha k}^\ast t_{\alpha q}^{\phantom{\ast}})}
{k-q}e^{i(k-q)d/2},\\
A_{\alpha k,\beta q}&=i
\frac{
(r_{\alpha k}^\ast t_{\beta q}^{\phantom{\ast}}
+t_{\alpha k}^\ast r_{\beta q}^{\phantom{\ast}})}
{k-q}e^{i(k-q)d/2},
\end{aligned}
\label{eq:ovlsca}
\end{equation}
where we have dropped terms suppressed by the small parameter $1/(d k_{\rs
F})$ and $\alpha\neq\beta$. For coinciding wave vectors $k=q$ the result reads
(again, $\alpha\neq \beta$)
\begin{equation}
\begin{aligned}
A_{\alpha k,\alpha k}&=-i
(r_{\alpha k}^\ast \partial^{\phantom{\ast}}_k
r_{\alpha k}^{\phantom{\ast}}
+t_{\alpha k}^\ast \partial^{\phantom{\ast}}_k
t_{\alpha k}^{\phantom{\ast}}),\\
A_{\alpha k,\beta k}&=-i
(r_{\alpha k}^\ast \partial^{\phantom{\ast}}_k
t_{\beta k}^{\phantom{\ast}}
+t_{\alpha k}^\ast \partial^{\phantom{\ast}}_k
r_{\beta k}^{\phantom{\ast}}).
\end{aligned}
\end{equation}
It is straightforward to generalize this scheme to situations where the kernel
$f(x \in [-d/2,d/2])=1$ decays smoothly outside the (scattering) region
$[-d/2,d/2]$. For the interacting resonant level model
\cite{vigman:78,schlottmann:78}, the interaction can be described by the
Hamiltonian Eq.\ (\ref{eq:Hint_s}) with a kernel $U(x,x') = U_0 g(x) f(x')$
and the function $g(x)$ finite in the regions nearby but outside the dot; the
interaction Hamiltonian Eq.\ (\ref{eq:Hint}) then can be expressed by the
product of matrix elements $A_{\alpha k,\beta q}$ and a similar quantity with
$f(x)$ replaced by $g(x)$ in Eq.\ (\ref{eq:ov1}).  In the following, we
concentrate on the capacitive interaction Eq.\ (\ref{eq:kernel}) with the
specific choice Eq.~(\ref{eq:f}).

Another situation where the knowledge of the scattering coefficients suffices
to evaluate matrix elements is the interaction with long wave-length bosonic
modes. As a specific example, we consider plasmonic excitations of a
two-dimensional electron gas (2DEG) induced by scattering electrons, e.g., at
a point contact; the interaction is described by the Hamiltonian
\cite{lebedev:10},
\begin{equation}
      \hat{H} = \frac1{2m}\sum_\sigma \int dx\,
      \hat\psi_\sigma^\dagger(x) \Bigl[ -i\hbar\partial_x -
      \frac{e}{c}\, \hat{A}(x) \Bigr]^2 \hat\psi_\sigma(x),
      \label{eq:boson}
\end{equation}
where $\hat{A}(x)$ is the vector potential of the electromagnetic field
induced by the two-dimensional plasmons. For the lowest transverse component
of the plasmon wave function $\phi_k(x)$ one can write,
\begin{equation}
      \hat{A}(x) = \sum_k ik\gamma \Bigl( \frac{2\pi\hbar
      c^2}{\omega_k} \Bigr)^{1/2}\bigl[ \hat{b}_k \phi_k(x) -
      \hat{b}_k^\dagger \phi_k^*(x) \bigr],
\end{equation}
where $\hat{b}_k^\dagger$ $(\hat{b}_k)$ are bosonic creation (annihilation)
operators for the plasmon modes with wave vector $k$ and linear dispersion
$\omega_k = v_\mathrm{pl}k$, with the plasmon velocity $v_\mathrm{pl}$ 
and a dimensionless geometrical factor $\gamma \sim 1$.

Typically, $v_\mathrm{pl} \gg v_{\rm\scriptscriptstyle F}$, and the
conservation of energy $\delta E = v_\mathrm{pl} \hbar k_\mathrm{pl} \gg
v_{\rm\scriptscriptstyle F} \hbar k$ implies that the plasmon wave length is
much larger than the Fermi wave length, $\lambda_\mathrm{pl} \gg
\lambda_{\rm\scriptscriptstyle F}$. The size of the plasmon then exceeds, by
several orders of magnitude, the scattering region (the size of the quantum
point contact) and thus the interaction between the plasmons and electrons is
dominated by the asymptotic scattering region where the electronic wave
function has the form~(\ref{eq:asymp}). Neglecting the spatial dependence of
the plasmon wave function $\phi(x) = \mathrm{const.}$, one again can express
the interaction Hamiltonian~(\ref{eq:boson}) solely in terms of the
non-interacting scattering matrix.

\subsection{Correction to Green's functions and scattering matrix}

We are interested in calculating the transport properties of a device, such as
the mean current and the current-noise correlator. Within the Green's function
approach, we need to know its single-particle version for the calculation
of the current, while for the calculation of the current-current correlator,
knowledge of the two-particle Green's function is required. Below, we first
derive the interaction correction to the single-particle and two-particle
Green's functions, before focusing on the transport properties in the next section.

\subsubsection{Single-particle Green's function}

The single-particle Keldysh Green's functions\cite{keldysh:65} are defined as
\begin{align}\label{eq:spg}
&G_{\nu\nu',\sigma}(x,t;x',t') \\ \nonumber
&\qquad=-i
\Bigl\langle
\mathcal{T}_K
\bigl[
\hat{\psi}^{\phantom{\dagger}}_{\sigma}(x,t_{\nu})
\hat{\psi}^{\dagger}_{\sigma}(x',t'_{\nu'})
\hat{S}_{K}
\bigr]
\Bigr\rangle,
\end{align}
where the expectation value of an operator $\hat{A}$ is defined as
$\langle\hat{A}\rangle=\text{Tr}\{\hat{A} \hat{\rho}_{\rs 0}\}$ with the
trace taken over all many-particle states of the system and $\hat{\rho}_{\rs
0}$ is given by Eq.~\eqref{eq:rho0}. We will skip the spin-index for the
single-particle Green's functions in the following, as these are equal
for both spin directions. The time-dependence of the field operator
$\hat{\psi}(x,t)$ in the interaction representation is governed by the free
Hamiltonian Eq.~(\ref{eq:H0LS}).  $\nu$ ($\nu'$) is a Keldysh index $\pm$
telling on which branch of the Keldysh contour $\mathcal{C}=(t_{\rs
0},\infty)_+\cup(\infty,t_{\rs 0})_-$ the time argument $t$ ($t'$) is
located (with $t_{0}\rightarrow-\infty$). $\mathcal{T}_K$ is the Keldysh
time-ordering operator and $\hat{S}_{K}$ is the Keldysh evolution operator
in the interaction representation. 

The non-interacting Green's functions in frequency representation
$G_{\nu\nu'}^{\rs (0)}(x,x',\omega)$ can be written in matrix
form\cite{keldysh:65}
\begin{align}
\label{eq:g0}
\hat{G}^{\rs (0)}(x,x',\omega)
=\frac{-i}{v_{\rm\scriptscriptstyle F}}
\sum_{\alpha}
&\biggl(
\varphi_{ \alpha k_{\omega}}^{\phantom{\ast}}(x)
\varphi_{ \alpha k_{\omega}}^{\ast}(x')
\hat{M}_{\alpha \omega}\\
&+\!\int\!\!
 \frac{dk}{2\pi i}
\frac{
\varphi_{\alpha k}^{\phantom{\ast}}(x)
\varphi_{\alpha k}^{\ast}(x')}
{k-k_{\omega} - i\delta}
\,\hat{\sigma}_{z}\biggr)\nonumber
\end{align}
with the wavevector $k_{\omega}=k_{\rm\scriptscriptstyle F} +(\omega-
\omega_{\rm\scriptscriptstyle F})/v_{\rm\scriptscriptstyle F}$, the matrix
\begin{align}
\hat{M}_{\alpha \omega}=
\begin{pmatrix}
-n_{\alpha}(\omega)  & -n_{\alpha}(\omega) \\
1-n_{\alpha}(\omega) & 1-n_{\alpha}(\omega)
\end{pmatrix},
\label{eq:Malphaomega}
\end{align}
and the Pauli matrix $\hat{\sigma}_{z}$. Here, $n_{\scriptscriptstyle
L/R}(\omega)$ is the Fermi distribution for the left/right lead. The integral
in the expression Eq.~(\ref{eq:g0}) of the non-interacting Green's functions
can be explicitly calculated for coordinates outside of the scattering
region, i.e., $|x|,|x'|>d/2$, by inserting the asymptotic wave functions
Eq.~(\ref{eq:asymp}). E.g., for $x>x'>d/2$, we obtain
\begin{align}
\label{eq:g0ev}
\hat{G}^{\rs (0)}(x,x',\omega)
=
&\phantom{+}\frac{-i}{v_{\rm\scriptscriptstyle F}}
e^{i k_\omega(x-x')}
(\hat{M}_{L\omega}+\hat{\sigma}_z)\\
&+\frac{-i}{v_{\rm\scriptscriptstyle F}}
e^{-i k_\omega (x-x')}
\hat{M}_{R\omega}\nonumber\\
&+\frac{-i}{v_{\rm\scriptscriptstyle F}}
R^{\phantom{\ast}}_{k_\omega}e^{i k_\omega (x-x')}
(\hat{M}_{R\omega}-\hat{M}_{L\omega})
\nonumber\\
&+\frac{-i}{v_{\rm\scriptscriptstyle F}}
r_{R k_\omega}^{\phantom{\ast}}e^{i k_\omega (x+x'-d)}
(\hat{M}_{R\omega}+\hat{\sigma}_z)\nonumber\\
&+\frac{-i}{v_{\rm\scriptscriptstyle F}}
r_{R k_\omega}^{\ast} e^{-i k_\omega(x+x'-d)}
\hat{M}_{R\omega}\nonumber
\end{align}
where $R^{\phantom{\ast}}_{k_{\omega}}=|r^{\phantom{\ast}}_{Rk_{\omega}}|^2$
and $T^{\phantom{\ast}}_{k_{\omega}}=|t^{\phantom{\ast}}_{Lk_{\omega}}|^2$
and we have used $T^{\phantom{\ast}}_{k_{\omega}} +
R^{\phantom{\ast}}_{k_{\omega}}=1$ to simplify the expression. Eq.\
(\ref{eq:g0ev}) expresses the non-interacting Green's functions outside
the scattering region in terms of scattering matrix coefficients; as we
will see below, the (lowest order) corrections due to interaction assume
the same structure and can be formulated in terms of a renormalization of
the scattering coefficients.
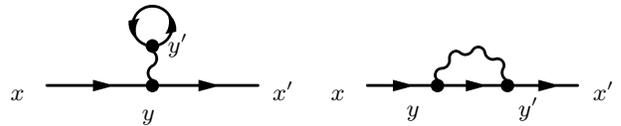
\begin{figure}[h]
\begin{center}
\begin{fmffile}{hartree}
  \fmfframe(1,2)(1,2){
\begin{fmfgraph*}(80,30)
    \fmfset{arrow_len}{3mm}
        \fmfbottom{i}
        \fmftop{v3}
        \fmfipair{l,b,r}
        \fmfiequ{l}{sw}
        \fmfiequ{b}{.5[sw,se]}
        \fmfiequ{r}{se}
        \fmfiv{lab=$x\phantom{'}$}{l}
        \fmfiv{lab=$x'$}{r}
        \fmfiv{lab=$y\phantom{'}$}{b}
        \fmfi{fermion}{l--b}
        \fmfi{fermion}{b--r}
        \fmf{photon}{i,v1}
        \fmflabel{$y'$}{v1}
        \fmf{fermion,right,tension=.5}{v1,v3}
        \fmf{fermion,right,tension=.5}{v3,v1}
        \fmfdot{i,v1}
\end{fmfgraph*}
}
\hspace{30pt}
  \fmfframe(1,2)(1,2){
\begin{fmfgraph*}(80,30)
    \fmfset{arrow_len}{3mm}
    \fmfstraight
        \fmfbottom{b1,b2,b3,b4}
        \fmflabel{$x\phantom{'}$}{b1}
        \fmflabel{$y\phantom{'}$}{b2}
        \fmflabel{$y'$}{b3}
        \fmflabel{$x'$}{b4}
        \fmf{fermion}{b1,b2}
        \fmf{fermion}{b2,b3}
        \fmf{fermion}{b3,b4}
        \fmf{photon, left}{b2,b3}
        \fmfdot{b2,b3}
\end{fmfgraph*}
}
\end{fmffile}
\end{center}
\caption{The Hartree and Fock diagrams give rise to the first order
corrections to the Green's functions given by Eq.~(\ref{eq:G1hh})
and~(\ref{eq:G1ff}).}
\label{fig:hf}
\end{figure}

The first-order correction $\hat{G}^{\rs (1)}$ to the single-particle Green's
function is given by the first-order expansion of Eq.~(\ref{eq:spg}) in
the interaction Hamiltonian.  There are two contributions in lowest-order,
a Hartree and a Fock term, cf.\ Fig.~\ref{fig:hf}, for which we find
the expressions
\begin{widetext}
\begin{align}
\hat{G}^{\rs (1H)}(x,x',\omega)
&=-\frac{2i}{\hbar}\int dy \,dy' \,U(y,y')
\int\!\frac{d\tilde{\omega}}{2\pi}\,
\hat{G}^{\rs (0)}(x,y,\omega)\hat{\sigma}_{z}
G_{\rs+-}^{\rs (0)}(y',y',\tilde{\omega})
\hat{G}^{\rs (0)}(y,x',\omega),\label{eq:G1hh}\\
\hat{G}^{\rs (1F)}(x,x',\omega)
&=\frac{i}{\hbar}\int dy \,dy' \,U(y,y')
\int\!\frac{d\tilde{\omega}}{2\pi}\,
\hat{G}^{\rs (0)}(x,y,\omega)\hat{\sigma}_{z}
G_{\rs+-}^{\rs (0)}(y,y',\tilde{\omega})
\hat{G}^{\rs (0)}(y',x',\omega).
\label{eq:G1ff}
\end{align}
\end{widetext}
The real-space non-interacting Green's functions $\hat{G}^{\rs (0)}(x,x',\omega)$ can be 
expressed through the scattering states using
\begin{align}
\hat{G}^{\rs (0)}(x,x',\omega)=\sum\limits_\alpha \int \frac{dk}{2\pi}
\hat{g}_{\alpha k}^{\rs (0)}(\omega)\varphi^{\phantom{\ast}}_{\alpha k}(x)\varphi^\ast_{\alpha k}(x'),
\end{align}
where $\hat{g}_{\alpha k}^{\rs (0)}(\omega)$ are the non-interacting Green's functions (in $k$-space)
\begin{align}
\hat{g}^{\rs (0)}_{\alpha k}(\omega)=g^{R}_{k}(\omega)
(\hat{M}_{\alpha \omega}+\hat{\sigma}_z)
-g^{A}_{k}(\omega)
\hat{M}_{\alpha \omega}
\label{eq:gk}
\end{align}
with the retarded/advanced Green's functions 
$g_{k}^{R/A}(\omega)=(\omega-\omega_k\pm i\delta)^{-1}$ and $\omega_k=E_k/\hbar$.
For our kernel Eq.~(\ref{eq:kernel}) and the specific choice of $f(x)$ in Eq.~(\ref{eq:f}), 
the integrations over the internal coordinates reduce to integrations of two scattering states
within the region $[-d/2,d/2]$. We replace these integrals by integrations over
the asymptotic regions as in Sec.~\ref{sec:ime} before in order to avoid the
integration over the scattering region. As a result, these overlaps can be expressed through
the matrix ele\-ments $A_{\alpha k,\beta q}$ defined in Eq.~(\ref{eq:ovlsca}). The expressions 
for the Hartree and Fock corrections can be evaluated for all coordinates outside of the
scattering region. Comparing for each choice of coordinate positions (to the
left/right of the scatterer) the corrections with the non-interacting Green's
function itself, we recognize that we can cast all effects of interaction into
a renormalization of the scattering coefficients. E.g., for $x,x'>d/2$, the
corrections assume the form of the three last terms in Eq.~(\ref{eq:g0ev}) and
we can define the renormalization $r^{\rs (1)}_{Rk}$ to the reflection
coefficient $r_{Rk}$ in a consistent manner such as to capture all three
corrections.  The corrections to the scattering matrix again involve Hartree-
and Fock contributions, $t^{\rs (1)}_{\alpha k}=t^{\rs (1H)}_{\alpha k}
+t^{\rs (1F)}_{ \alpha k}$ and $r^{\rs (1)}_{\alpha k}= r^{\rs (1H)}_{\alpha
k}+r^{\rs (1F)}_{\alpha k}$; the Hartree terms are given by
\begin{align}\label{eq:scaHartree}
t^{\rs (1H)}_{Lk}&=-\frac{2U_{\rs 0}}{\hbar v_{\rm\scriptscriptstyle F}}\sum_{\beta}
\int \frac{dq}{2\pi} n_{\beta q}
\bigl(it_{k}\, A_{Lk,Lk} A_{\beta q,\beta q} \\ \nonumber
&\hspace{100pt}+ir_{Rk}\, A_{Rk,Lk} A_{\beta q,\beta q} \bigr),\\ \nonumber
r^{\rs (1H)}_{Rk}&=-\frac{2U_{\rs 0}}{\hbar v_{\rm\scriptscriptstyle F}}\sum_{\beta}
\int \frac{dq}{2\pi} n_{\beta q}
\bigl(ir_{Rk}\, A_{Rk,Rk}A_{\beta q,\beta q} \\ \nonumber
&\hspace{100pt}+it_{k}\, A_{Lk,Rk}A_{\beta q,\beta q}\bigr),
\end{align}
and the corresponding corrections $t^{\rs (1H)}_{Rk}$ and $r^{\rs (1H)}_{Lk}$
are obtained by interchanging ${\rm R}\leftrightarrow {\rm L}$. The $q$- and
$k$-dependent parts in the above expressions factorize and we find the 
simpler results 
\begin{equation}
\begin{aligned}
t^{\rs (1H)}_{\alpha k}&=
-\frac{U_{\rs 0}N_{\rm int}}{\hbar v_{\rm\scriptscriptstyle F}}
\partial^{\phantom{\ast}}_k t_{k},\\
r^{\rs (1H)}_{\alpha k}&=
-\frac{U_{\rs 0}N_{\rm int}}{\hbar v_{\rm\scriptscriptstyle F}}
\partial^{\phantom{\ast}}_k r_{\alpha k},
\end{aligned}
\label{eq:tr1H}
\end{equation}
with the particle number $N_{\rm int}$ in the interaction region given by
\begin{align}\label{eq:Nint}
N_{\rm int}&=
\sum_\sigma
\int\limits_{-d/2}^{d/2}\!\! dy\,
\langle
\hat{\psi}^\dagger_\sigma(y,t)
\hat{\psi}^{\phantom{\dagger}}_\sigma(y,t)
\rangle\nonumber\\
&=2 \int \frac{dq}{2\pi}
\bigl(n_{ L q}A_{L q,L q}
+n_{ R q}A_{ R q,R q}\bigr),
\end{align}
where $n_{\alpha k}=n_{\alpha}(E_k)$ and the factor 2 accounts for the two
spin directions (note that the $k$-dependent factors in Eq.\
(\ref{eq:scaHartree}) reduce to a derivative of the scattering amplitudes,
i.e., the brackets in Eq.\ (\ref{eq:scaHartree}) reduce to $A_{\beta q,\beta
q} \partial_k t_{k}$ and $A_{\beta q,\beta q} \partial_k r_{Rk}$,
respectively).

Given the corrections to the scattering amplitudes, we can calculate the
corrections to the transmission and reflection probabilities $T^{\rs
(1H)}_{k}$ and $R^{\rs (1H)}_{k}=-T^{\rs (1H)}_{k}$, 
\begin{align}
T^{\rs (1H)}_{k}&=-\frac{U_{\rs 0}N_{\rm int} }{\hbar v_{\rm\scriptscriptstyle F}}
\partial^{\phantom{\ast}}_{k} T_{k}.
\label{eq:T1H}
\end{align}
The result is easily interpreted: the Hartree interaction effectively shifts
the scattering potential with respect to the energy of the incoming particle
and thus the transmission characteristics of the scatterer is shifted
accordingly; Eq.~(\ref{eq:T1H}) then describes the lowest order correction of
$T_{k}$ due to a shift in wavelength by $\delta k=-U_{\rs 0}N_{\rm int}/\hbar
v_{\rm\scriptscriptstyle F}$.

The Fock correction to the scattering matrix assumes the form
\begin{align}\label{eq:scaFock}
t^{\rs (1F)}_{Lk}&=\frac{U_{\rs 0}}{\hbar v_{\rm\scriptscriptstyle F}}\sum_{\beta}
\int \frac{dq}{2\pi} n_{\beta q}
\bigl(it_{k}\, A_{Lk,\beta q} A_{\beta q,Lk} \\ \nonumber
&\hspace{90pt}+ir_{Rk}\, A_{Rk,\beta q} A_{\beta q,Lk} \bigr),\\ \nonumber
r^{\rs (1F)}_{Rk}&=\frac{U_{\rs 0}}{\hbar v_{\rm\scriptscriptstyle F}}\sum_{\beta}
\int \frac{dq}{2\pi} n_{\beta q}
\bigl(ir_{Rk}\, A_{Rk,\beta q}A_{\beta q,Rk} \\ \nonumber
&\hspace{90pt}+it_{k}\, A_{Lk, \beta q}A_{\beta q,Rk}\bigr).
\end{align}
The corrections $t^{\rs (1F)}_{R k}$ and $r^{\rs (1F)}_{L k}$ can be found
by exchanging $\text{R}\leftrightarrow \text{L}$ in the expressions above.
While for the Hartree term the corrections to the transmission
coefficients are equal, i.e., $t_{Rk}^{\rs (1H)}=t_{Lk}^{\rs (1H)}$, this is
in general not the case for the Fock contribution (however, equality holds
in equilibrium).  The corrections to the transmission and reflection
probabilities $T^{\rs (1F)}_{k}=-R^{\rs (1F)}_{k}$ are
\begin{align}
\label{eq:T1F}
&T^{\rs (1F)}_{k}=-\frac{U_{\rs 0}}{\hbar v_{\rm\scriptscriptstyle F}}
\sum_{\beta}\int \frac{dq}{2\pi} \frac{n_{\beta q}}{(k-q)^2}\\
&\hspace{15pt}\times\Bigl[
i r_{\beta k}^{\phantom{\ast}} t_{k}^{\ast}
(t_{k}^{\phantom{\ast}} r_{\beta q}^{\ast}
+t_{q}^{\phantom{\ast}} r_{\beta k}^{\ast}
-t_{q}^{\phantom{\ast}} r_{\beta q}^{\ast})
\,\,+\,\,\text{c.c.}\Bigr].\nonumber
\end{align}
The modifications to the current and the current-current correlator due to the
renormalization of the single-particle Green's functions will be discussed in
section~\ref{sec:c} below. Higher order perturbation theory might generate
terms of different form which cannot be incorporated within a simple
renormalization of the scattering coefficients.

\subsubsection{Two-particle Green's function}
\label{sec:tpg}

The two-particle Keldysh Green's functions are defined as
\begin{align}\label{eq:tpg}
&\mathcal{G}_{\sigma\pi}(1,2;1',2')\\
\nonumber
&\qquad=
(-i)^2
\Bigl\langle
\mathcal{T}_K
\bigl(
\hat{\psi}^{\phantom{\dagger}}_{\rs \sigma}(1)
\hat{\psi}^{\phantom{\dagger}}_{\rs \pi}(2)
\hat{\psi}^{\dagger}_{\rs \pi}(2')
\hat{\psi}^{\dagger}_{\rs \sigma}(1')
\hat{S}_{K}
\bigr)
\Bigr\rangle,
\end{align}
where the variables $i=1,2,1',2'$ represent both space and time coordinates
as well as the Keldysh index, i.e., $i=(x_i,t_i,\nu_i)$.  The two-particle
Green's functions can be split into a reducible and an irreducible
part, $\mathcal{G}_{\sigma\pi}=\mathcal{G}^{\rs(red)}_{\sigma\pi}
+\mathcal{G}^{\rs(irr)}_{\sigma\pi}$, where the reducible part can be
expressed through single-particle Green's functions as
\begin{align} \label{eq:tpgr}
\mathcal{G}^{{\rs (red)}}_{\sigma\pi} (1,2;1',2')
&=G(1,1') G(2,2')\\ \nonumber
&\qquad-\delta_{\sigma\pi}G(1,2') G(2,1')
\end{align}
with the single-particle Green's function $G(1,1')=G_{\nu_1\nu'_{1}}
(x_1,t_1,x'_{1},t'_{1})$, cf.\ Eq.\ (\ref{eq:spg}), and similar for the other
coordinates (again, we ignore the spin index).  The non-interacting reducible
two-particle Green's functions $(\mathcal{G}^{{\rs (red)}}_{\sigma\pi})^{\rs
(0)}$ and corrections due to interaction (e.g., $(\mathcal{G}^{{\rs
(red)}}_{\sigma\pi})^{\rs (1)}$) can be calculated from the non-interacting
single-particle Green's functions $\hat{G}^{\rs (0)}$ and the correction
$\hat{G}^{\rs (1)}$ to the single-particle Green's function.\\

\begin{figure}[h]
    \begin{center}
\begin{fmffile}{g2irr}
  \fmfframe(1,2)(1,2){
\begin{fmfgraph*}(60,40)
    \fmfset{arrow_len}{3mm} \fmfstraight \fmfleft{i1,i2}
    \fmfright{o1,o2} \fmf{fermion}{i1,v1,o1} \fmf{fermion}{i2,v2,o2}
    \fmfdot{v1,v2} \fmf{photon}{v1,v2} \fmflabel{$1\phantom{'}$}{i1}
    \fmflabel{$2\phantom{'}$}{i2} \fmflabel{$1^\prime$}{o1} \fmflabel{$2^\prime$}{o2}
    \fmflabel{$y\phantom{'}$}{v1} \fmflabel{$y^\prime$}{v2}
\end{fmfgraph*} } \hspace{40pt}
  \fmfframe(1,2)(1,2){
\begin{fmfgraph*}(60,40)
    \fmfset{arrow_len}{3mm} \fmfstraight \fmfleft{i1,i2}
    \fmfright{o1,o2} \fmf{fermion}{i1,v1} \fmf{phantom}{v1,o1}
    \fmf{fermion}{i2,v2} \fmf{phantom}{v2,o2} \fmf{photon}{v1,v2}
    \fmffreeze \fmf{fermion}{v1,o2} \fmf{fermion}{v2,o1} \fmfdot{v1,v2}
    \fmflabel{$1\phantom{'}$}{i1} \fmflabel{$2\phantom{'}$}{i2}
    \fmflabel{$1^\prime$}{o1} \fmflabel{$2^\prime$}{o2} \fmflabel{$y\phantom{'}$}{v1}
    \fmflabel{$y^\prime$}{v2}
\end{fmfgraph*} }
\end{fmffile}
\end{center}
\caption{
There are two processes contributing to the correction to the
irreducible two-particle Green's function in first order given by
Eq.~(\ref{eq:tpgi1}). While the second process only contributes for equal
spins, the first one also gives a contribution for opposite spins.
}
\label{fig:g2irr}
\end{figure}
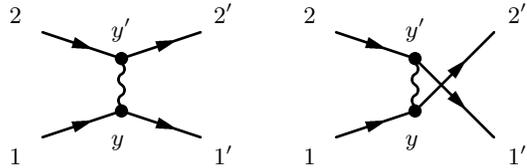

The remaining irreducible part $\mathcal{G}^{\rs(irr)}_{\sigma\pi}$ cannot be
expressed through single-particle Green's functions and is given by the
connected diagrams of Eq.~\eqref{eq:tpg}.  The lowest-order contribution to
the irreducible two-particle Green's function involves two contributions, cf.\
Fig.~\ref{fig:g2irr}, and can be evaluated for coordinates outside the
scattering region and arbitrary Keldysh indices using the non-interacting
Green's function in Eq.~(\ref{eq:g0}); to first order in the interaction
(note that $\bigl(\mathcal{G}^{\rs(irr)}_{\sigma\pi}\bigr)^{\rs (0)} = 0$), we
obtain in frequency representation with $\tilde{i}=(x_i,\omega_i,\nu_i)$,
\begin{widetext}
\begin{equation}
\begin{aligned}
&\bigl(\mathcal{G}^{\rs(irr)}_{\sigma\pi}\bigr)^{\rs (1)}
(\tilde{1},\tilde{2};\tilde{1}',\tilde{2}')
=i\frac{U_{\rs 0}}{\hbar}
2\pi \delta(\omega_1+\omega_2-\omega_1'-\omega_2')
\int\limits_{-d/2}^{d/2} \!\! dy \,
\, dy'\,
\\
&\hspace{120pt}\times\sum_{\gamma}
\gamma
\Bigl[
G^{\rs (0)}_{\s \nu_1 \gamma}(x_1,y,\omega_1)
G^{\rs (0)}_{\s \nu_2 \gamma}(x_2,y',\omega_2)
G^{\rs (0)}_{\s \gamma \nu_{2'}}(y',x_2',\omega_2')
G^{\rs (0)}_{\s \gamma \nu_{1'}}(y,x_1',\omega_1')\\
&\hspace{150pt}
-\delta_{\sigma \pi}
G^{\rs (0)}_{\s \nu_1 \gamma}(x_1,y,\omega_1)
G^{\rs (0)}_{\s \nu_2 \gamma}(x_2,y',\omega_2)
G^{\rs (0)}_{\s \gamma \nu_{2'}}(y,x_2',\omega_2')
G^{\rs (0)}_{\s \gamma \nu_{1'}}(y',x_1',\omega_1')
\Bigr].
\end{aligned}
\label{eq:tpgi1}
\end{equation}
\end{widetext}
The integrals over the region $[-d/2,d/2]$ are calculated in the same way as
in Sec.~\ref{sec:ime} above. 

\section{Correction to current and noise}\label{sec:c}

Next, we determine the interaction corrections to the current and
current-current correlator as calculated from single-particle and two-particle
Green's functions; the irreducible contributions to the latter will give rise
to spin correlations in an asymmetric device.

%
\subsection{Current}

The current $j(x,t)$ derives from the single-particle Green's function via
\begin{align}
j(x,t)&= 2 \hat{\mathcal{D}}_x G_{\rs +-}(x,t,x',t),
\label{eq:jG}
\end{align}
where $\hat{\mathcal{D}}_x =(e\hbar/2m)(\partial_{x'}-\partial_x)$ and
the limit $x'\rightarrow x$ is taken in the end. The factor 2 accounts
for the two spin orientations of the electrons. We determine the current
to the right of the scattering and interaction region, $x>d/2$. As the
effect of interactions on the single-particle Green's function can be
expressed through a renormalization of the scattering coefficients, the
same applies to the current and we arrive at the well-known Landauer formula
\begin{align}
j^{(0)} + j^{(1)}=
2ev_{\rm\scriptscriptstyle F} \int \frac{dk}{2\pi} [T_{k} + T^{\rs (1)}_{k}]
(n_{L k}-n_{R k})
\label{eq:j0j1}
\end{align}
with $T_{k}^{\rs (1)}=T_{k}^{\rs (1H)}+T_{k}^{\rs (1F)}$ given in Eqs.\
(\ref{eq:T1H}) and (\ref{eq:T1F}).
%
\subsection{Current-current correlator}
%
Let us now turn our attention to the current-current correlator
\begin{align}\label{eq:S}
&S_{\sigma\pi}(\omega,x_1,x_2) \\
\nonumber
&\qquad=\int d\tau
\langle\langle
\mathcal{T}_{K}\bigl(
\hat{j}_{\sigma}(x_1,\tau_{-})
\hat{j}_{\pi}(x_2,0_{+})
\hat{S}_K \bigr)
\rangle\rangle e^{i\omega \tau}
\end{align}
behind the scatterer, $x_1,x_2>d/2$. Below, we focus on the zero-frequency
correlator which is independent of $x_1$ and $x_2$, i.e., $S_{\sigma\pi}(0)=
S_{\sigma\pi} (0,x_1,x_2)$, as explained in Ref.~\onlinecite{lesovik:99}.
The current-current correlator can be expressed in
terms of two-particle Green's functions and thus can be split into reducible
and irreducible parts, $S_{\sigma\pi}= S_{\sigma\pi}^{\rs(red)}
+S_{\sigma\pi}^{\rs(irr)}$. The reducible part vanishes for opposite spins
$\sigma\neq\pi$, while for equal spins it is expressible in terms of
single-particle Green's functions,
\begin{align}
    S_{\sigma\sigma}^{\rs (red)}(0)
    =-\hat{\mathcal{D}}_{x_1}\hat{\mathcal{D}}_{x_2}
    \int\frac{d\omega}{2\pi}
    G_{\rs -+}(x_1,x_2',\omega)
    G_{\rs +-}(x_2,x_1',\omega).
\label{eq:ccri}
\end{align}
The irreducible part contributies to both, equal and opposite spins, and can
be expressed as
\begin{align}
S_{\sigma\pi}^{\rs (irr)}(0)
=\hat{\mathcal{D}}_{x_1}\hat{\mathcal{D}}_{x_2}
\int\frac{d\omega_1\,d\omega_2}{(2\pi)^2}
\mathcal{G}^{\rs(irr)}_{\sigma\pi}(1_{\rs -},2_{\rs +},1'_{\rs -},2'_{\rs +}),
\label{eq:ccirr}
\end{align}
where we explicitely indicated the Keldysh indices and $\omega_{1',2'}
= \omega_{1,2}$.

In the non-interacting case, the irreducible Green's function
vanishes and the current-current correlator is given by the well-known
expression\cite{lesovik:89,buettiker:90}
\begin{align}
\label{eq:cc0}
&S^{\s (0)}_{\sigma\pi}(0)=\delta_{\sigma\pi}e^2 v_{\rm\scriptscriptstyle F}
\int\frac{dk}{2\pi}\\
&\hspace{25pt}\times\Bigl[T_{k}^{2} \,\,
\bigl( n_{Lk} (1-n_{Lk})+n_{Rk}(1-n_{Rk}) \bigr)\nonumber\\
&\hspace{35pt}+R_{k}T_{k}
\bigl( n_{Lk}(1-n_{Rk})+n_{Rk}(1-n_{Lk}) \bigr)
\Bigr].\nonumber
\end{align}
There are two kinds of (first-order) interaction corrections to the noise
correlator: those due to the reducible part $S_{\sigma\pi}^{\rs (red)}$ can be
expressed through the renormalization of the transmission probability $T_{k}$
in Eq.~(\ref{eq:cc0}). The contribution to the irreducible part
$S_{\sigma\pi}^{\rs (irr)}$ gives rise to additional correlations which go
beyond a simple renormalization.

\subsubsection{Reducible corrections}

The expansion of the reducible part Eq.~(\ref{eq:ccri}) in the interaction
produces corrections originating from those in the single-particle Green's
functions and we find
\begin{align}
&S_{\sigma\pi}^{\rs (red)(1)}(0)
=\delta_{\sigma \pi} e^2 v_{\rm\scriptscriptstyle F}\!\!\int\!\! \frac{dk}{2\pi}\\
&\hspace{15pt}\times\Bigl[2T_{k}T_{k}^{\rs (1)}
\bigl( n_{Lk}(1-n_{Lk})+n_{Rk}(1-n_{Rk}) \bigr)\nonumber\\
&\hspace{25pt}+(1-2 T_{k})T_{k}^{{\rs (1)}}
\bigl( n_{Lk}(1-n_{Rk})+n_{Rk}(1-n_{Lk}) \bigr)\Bigr],\nonumber
\end{align}
which corresponds to the non-interacting expression Eq.~(\ref{eq:cc0})
with a renormalized transmission probability $T_{k} + T_{k}^{\rs (1)}$
and expanded to first order in $T_{k}^{\rs (1)}$.
%
\subsubsection{Irreducible corrections}\label{sec:irred_corr}
%
Restricting ourselves to zero-temperature ($\vartheta=0$), we find the
first-order contribution to the irreducible two-particle Green's function
Eq.~(\ref{eq:tpgi1}) with all coordinates to the right of the barrier
$x_i,x_i'>d/2$ and evaluate the expression Eq.\ (\ref{eq:ccirr}) for the
irreducible noise correlator to arrive at
\begin{align}
\label{eq:cc2p}
&S_{\sigma\pi}^{\rs (irr)(1)}(0)\bigr|_{\vartheta=0}=
-e^2 \frac{U_{0}}{\hbar}
\text{sign}(\mu_{\rs L}-\mu_{\rs R})
\!\!\int\limits_{k_{\rs min}}^{k_{\rs max}}
\frac{dk_1\,dk_2}{(2\pi)^2}\hspace{25pt}\,
\nonumber\\
&\hspace{5pt}\times\Bigl[ \Bigl(i
r_{Lk_1}^\ast t_{k_1}^{\phantom{\ast}}
r_{Lk_2}^\ast t_{k_2}^{\phantom{\ast}}
\, A_{Rk_1,Lk_1}A_{Rk_2,Lk_2}\nonumber\\
&\hspace{25pt}
-i
r_{Rk_1}^\ast t_{k_1}^{\phantom{\ast}}
r_{Rk_2}^\ast t_{k_2}^{\phantom{\ast}}
\, A_{Lk_1,Rk_1}A_{Lk_2,Rk_2}
\Bigr)\nonumber\\
&\hspace{15pt} -\delta_{\sigma \pi} \Bigl(i
r_{Lk_1}^\ast t_{k_1}^{\phantom{\ast}}
r_{Lk_2}^\ast t_{k_2}^{\phantom{\ast}}
\, A_{Rk_1,Lk_2}A_{Rk_2,Lk_1}\nonumber\\
&\hspace{40pt}
-i
r_{Rk_1}^\ast t_{k_1}^{\phantom{\ast}}
r_{Rk_2}^\ast t_{k_2}^{\phantom{\ast}}
\, A_{Lk_1,Rk_2}A_{Lk_2,Rk_1}
\Bigr)\Bigr],
\end{align}
where $k_{\rs max/min}=\text{max/min}(k_{\rm\scriptscriptstyle FL} /
k_{\rm\scriptscriptstyle FR})$. The above correction vanishes for the
symmetric setup where $r_{Lk} = r_{Rk}$ and $A_{Rk,Lq}=A_{Lk,Rq}$; on the
other hand, interactions combined with an asymmetric scatterer, a case that
has not been addressed previously, generate a finite correlation already in
lowest order in $U_0$.

\begin{figure}
\includegraphics{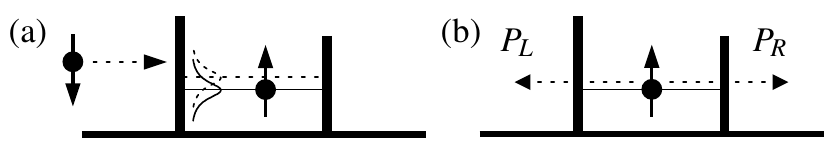}
\caption{(a) A (spin-up) electron on the dot shifts the resonance
level for incoming (spin-down) electrons, thereby modulating the (spin-down)
current across the scatterer.  (b) For an asymmetric scatterer, the escape
probability of the (spin-up) electron to the left ($P_L$) and right
($P_R$) leads are different, generating a net (spin-up) current which is
correlated with the (spin-down) current modulation.}
\label{fig:asymm}
\end{figure}

In summary, for {\it equal spins}, the correlation (or noise), due to
scattering and fermionic statistics already present without interaction, is
modified due to interactions via the renormalization of the transmission (in
the reducible contribution) and by the additional irreducible contribution
given by Eq.~(\ref{eq:cc2p}) (note that this first-order contribution may
vanish for equal spins due to the Pauli principle, as, e.g., for a
single-level quantum dot).  {\it Opposite-spin} correlations are absent in the
non-interacting case, and a finite result is generated to first order in the
interaction for an asymmetric scatterer by the first pair of terms in
Eq.~(\ref{eq:cc2p}) and to second order in $U_0$ for a symmetric scatterer.
We attribute the origin of the asymmetry-induced correlations to the following
effect (we restrict the discussion to the case of an asymmetric quantum dot,
cf.\ Fig.\ \ref{fig:asymm}): First of all, we note that, by virtue of Pauli's
exclusion principle, all interaction-induced modifications of the noise
correlator are due to the presence of spin $\sigma$ particles on the dot
modifying the current of spin $-\sigma$ particles traversing the dot.  The
finite correlation arises from the change in current $\delta I_{-\sigma}$ due
to the presence of particles $\delta N_{\mathrm{int}\sigma}$ on the dot, cf.\
Fig.~\ref{fig:asymm} (a).  For a symmetric scatterer, the decay of the charge
$\delta N_{\mathrm{int}\sigma}$ to the unblocked scattering state is symmetric
and the undirected current leads to a vanishing current-current correlator. On
the other hand, in an asymmetric dot, the charge $\delta N_{\mathrm{int}
\sigma}$ decays in an asymmetric manner and its current $\delta I_\sigma$ is
correlated with $\delta I_{-\sigma}$, cf.\ Fig.~\ref{fig:asymm} (b).
%

\section{Single resonance level}\label{sec:srl}

The above results, which apply to an arbitrary scatterer, can be simplified and
further developed when choosing a particular scatterer. Here, we consider a
simple example in the form of a scatterer with a single resonance level at
energy $E_{\rm res}=E_{\rm\scriptscriptstyle F}+\hbar v_{\rm\scriptscriptstyle
F} (k_{\rm res}-k_{\rm\scriptscriptstyle F})$, located at $x=0$ with vanishing
extent $d\rightarrow 0$. This single-level quantum dot is described by the 
scattering matrix $S_{k}$ with amplitudes
\begin{align}
t_k&=\frac{i\gamma}{\Delta k+i\gamma}\sqrt{1-\eta^2},
    \nonumber\\
r_{L k}&=\frac{\Delta k+i\eta\gamma}
    {\Delta k+i\gamma},
    \\
r_{R k}&=
    \frac{\Delta k-i\eta\gamma}
    {\Delta k+i\gamma},
    \nonumber
\end{align}
where the wavevector $\Delta k=k-k_{\rs res}$ is measured relative to the
resonance level; the level width $\hbar v_{\rm\scriptscriptstyle F}\gamma$
is parametrized by $\gamma$ and the dimensionless parameter $\eta \in
[-1,1]$ characterizes the asymmetry of the system. This scattering matrix
generates a transmission probability $T_k=|t_k|^2=(1-\eta^2)/(1+(\Delta
k/\gamma)^2)$.

The scattering matrix of the single resonance level model gives rise to
matrix elements $A_{\alpha k, \beta q}$ of a particularly simple
form, i.e., 
\begin{align}
A_{\alpha k,\beta q}=a_{\alpha}^\ast \phi_{k}^\ast  
a_{\beta}^{\phantom{\ast}}\phi_{q}^{\phantom{\ast}},
\label{eq:Asrlm}
\end{align}
with $a_{L/R}=\pm i\sqrt{1\mp \eta}$ and $\phi^{\phantom{\ast}}_k=\sqrt{\gamma}
/(\Delta k+i\gamma)$. The interaction Hamiltonian Eq.\ (\ref{eq:Hint}) then
factorizes in the momenta, with individual terms given by
\begin{align}\label{eq:hint}
\hat{h}_{\rm int}&=\frac{U_{0}}{2} \,
a^\ast_{\alpha_1'}a^\ast_{\alpha_2'}
a^{\phantom{\ast}}_{\alpha_2}a^{\phantom{\ast}}_{\alpha_1}\,
\phi_{k_1'}^\ast \phi_{k_2'}^\ast
\phi_{k_2}^{\phantom{\ast}}\phi_{k_1}^{\phantom{\ast}} \\
&\hspace{40pt}
\times \delta_{\sigma_1\sigma_1'}\delta_{\sigma_2\sigma_2'}\,
\hat{c}_{\alpha_1' k_1' \sigma_1'}^\dagger
\hat{c}_{\alpha_2' k_2' \sigma_2'}^\dagger
\hat{c}_{\alpha_2 k_2 \sigma_2}^{\phantom{\dagger}}
\hat{c}_{\alpha_1 k_1 \sigma_1}^{\phantom{\dagger}},\nonumber
\end{align}
and integration, summation over all wave vectors, lead, and spin indices
providing the full interaction Hamiltonian $\hat{H}_{\rm int}$.  Introducing
the anti-symmetrized two-particle vertex 
\begin{align}
\Gamma^{(0)}_{1'2',2 1}=U_{0}
a^\ast_{\alpha_1'}a^\ast_{\alpha_2'}
a^{\phantom{\ast}}_{\alpha_2}a^{\phantom{\ast}}_{\alpha_1}
\phi_{k_1'}^\ast \phi_{k_2'}^\ast
\phi_{k_2}^{\phantom{\ast}}\phi_{k_1}^{\phantom{\ast}}
S_{\sigma_1'\sigma_2',\sigma_2\sigma_1}
\label{eq:Gamma0}
\end{align}
with $i=1,~2,~1',~2'$ representing the multi-index $(\alpha_i,k_i,\sigma_i)$ and
\begin{align} 
S_{\sigma_1'\sigma_2',\sigma_2\sigma_1}=\delta_{\sigma_1'\sigma_1}
\delta_{\sigma_2'\sigma_2}-\delta_{\sigma_1'\sigma_2}
\delta_{\sigma_2'\sigma_1},
\end{align}
the individual terms in Eq.\ (\ref{eq:hint}) assume the form
\begin{align}
\hat{h}_{\rm int}=\frac{1}{4}\Gamma^{(0)}_{1'2',2 1}
\hat{c}_{\alpha_1' k_1' \sigma_1'}^\dagger
\hat{c}_{\alpha_2' k_2' \sigma_2'}^\dagger
\hat{c}_{\alpha_2 k_2 \sigma_2}^{\phantom{\dagger}}
\hat{c}_{\alpha_1 k_1 \sigma_1}^{\phantom{\dagger}}.
\end{align}
The Hartree and Fock digrams in Fig.\ \ref{fig:hf} then collapse into one
diagram, cf.\ Fig.\ \ref{fig:vertex}.
\begin{figure}
\begin{tabular}{p{20pt}m{50pt}m{30pt}m{25pt}m{30pt}m{25pt}m{30pt}}
(a)\\ &
$\Gamma^{(0)}_{1^\prime 2^\prime,21}=$ \hspace{5pt}&
\begin{fmffile}{Gamma}
  \fmfframe(1,2)(1,2){
\begin{fmfgraph*}(20,20)
    \fmfset{arrow_len}{2mm} \fmfstraight \fmfleft{i1,i2}
    \fmfright{o1,o2} 
    \fmf{dots_arrow,tension=1.2}{i1,p1}
    \fmf{dots,tension=1.2}{p1,v1}
    \fmf{dots,tension=1.2}{w1,q1} 
    \fmf{dots_arrow,tension=1.2}{q1,o1} 
    \fmf{dots_arrow,tension=1.2}{i2,p2}
    \fmf{dots,tension=1.2}{p2,v2}
    \fmf{dots,tension=1.2}{w2,q2}
    \fmf{dots_arrow,tension=1.2}{q2,o2}
    \fmf{plain,tension=.8}{v1,w1}
    \fmf{plain,tension=.8}{v2,w2}
    \fmf{plain,tension=.8}{v1,v2}
    \fmf{plain,tension=.8}{w1,w2}
    \fmflabel{$1\phantom{'}$}{i1}
    \fmflabel{$2\phantom{'}$}{i2} \fmflabel{$1^\prime$}{o1} 
    \fmflabel{$2^\prime$}{o2}
\end{fmfgraph*} }
\end{fmffile}&
=
&
\begin{fmffile}{Gammaeqop}
  \fmfframe(1,2)(1,2){
\begin{fmfgraph*}(20,20)
    \fmfset{arrow_len}{2mm} \fmfstraight \fmfleft{i1,i2}
    \fmfright{o1,o2} 
    \fmf{dots_arrow,tension=1.2}{i1,p1}
    \fmf{dots,tension=1.2}{p1,v1}
    \fmf{dots,tension=1.2}{v1,q1} 
    \fmf{dots_arrow,tension=1.2}{q1,o1} 
    \fmf{dots_arrow,tension=1.2}{i2,p2}
    \fmf{dots,tension=1.2}{p2,v2}
    \fmf{dots,tension=1.2}{v2,q2}
    \fmf{dots_arrow,tension=1.2}{q2,o2}
    \fmf{photon,tension=.8}{v1,v2}
    \fmflabel{$1\phantom{'}$}{i1}
    \fmflabel{$2\phantom{'}$}{i2} \fmflabel{$1^\prime$}{o1} 
    \fmflabel{$2^\prime$}{o2}
\end{fmfgraph*} }
\end{fmffile}&
\hspace{5pt}-
&
\begin{fmffile}{Gammaop}
  \fmfframe(1,2)(1,2){
\begin{fmfgraph*}(20,20)
    \fmfset{arrow_len}{2mm} \fmfstraight \fmfleft{i1,i2}
    \fmfright{o1,o2} 
    \fmf{dots_arrow,tension=1.2}{i1,p1}
    \fmf{dots,tension=1.2}{p1,v1}
    \fmf{dots,tension=1.2}{v1,q1} 
    \fmf{dots_arrow,tension=1.2}{q1,o1} 
    \fmf{dots_arrow,tension=1.2}{i2,p2}
    \fmf{dots,tension=1.2}{p2,v2}
    \fmf{dots,tension=1.2}{v2,q2}
    \fmf{dots_arrow,tension=1.2}{q2,o2}
    \fmf{photon,tension=.8}{v1,v2}
    \fmflabel{$1\phantom{'}$}{i1}
    \fmflabel{$2\phantom{'}$}{i2} \fmflabel{$2^\prime$}{o1} 
    \fmflabel{$1^\prime$}{o2}
\end{fmfgraph*} }
\end{fmffile}\\
&&&&&\\
&&&&&\\
(b)\\ &
$\Sigma^{(1)}(\omega)=$&
\begin{fmffile}{self}
  \fmfframe(1,2)(1,2){
\begin{fmfgraph*}(20,20)
    \fmfset{arrow_len}{3mm}
	\fmfbottomn{b}{4}
	\fmftopn{t}{4}
	\fmf{phantom}{t2,v2}
	\fmf{phantom}{v2,w2}
	\fmf{plain}{w2,b2}
	\fmf{phantom}{t3,v3}
	\fmf{phantom}{v3,w3}
	\fmf{plain}{w3,b3}
	\fmf{phantom}{b1,b2}
    \fmf{phantom}{b3,b4}
    \fmf{plain,tension=0}{w2,w3}
    \fmf{plain}{b2,b3}
    \fmf{phantom}{t2,m,t3}
    \fmffreeze
    \fmf{phantom}{w2,n,w3}
    \fmffreeze
    \fmf{fermion,right}{n,m,n}
\end{fmfgraph*}
}
\end{fmffile}&
\hspace{5pt}=
&
\begin{fmffile}{selfhar}
  \fmfframe(1,2)(1,2){
\begin{fmfgraph*}(20,20)
    \fmfset{arrow_len}{3mm}
        \fmfbottom{i}
        \fmftop{v3}
        \fmfipair{b}
        \fmfiequ{b}{.5[sw,se]}
        \fmf{photon}{i,v1}
        \fmf{fermion,right,tension=.5}{v1,v3}
        \fmf{fermion,right,tension=.5}{v3,v1}
	\fmfdot{i,v1}
\end{fmfgraph*}
}
\end{fmffile}&
+
&
\begin{fmffile}{selffock}
  \fmfframe(1,2)(1,2){
\begin{fmfgraph*}(20,20)
    \fmfset{arrow_len}{3mm}
    \fmfstraight
        \fmfbottom{b1,b2}
        \fmf{fermion}{b1,b2}
        \fmf{photon, left}{b1,b2}
	\fmfdot{b1,b2}
\end{fmfgraph*}
}
\end{fmffile}
\end{tabular}
\caption{
(a) The two-particle vertex $\Gamma^{(0)}_{1^\prime 2^\prime,21}$ is 
anti-symmetrized with respect to the two incoming legs and to the two outgoing legs. 
(b) The first-order self-energy $\Sigma^{(1)}(\omega)$ consists of a Hartree and 
a Fock contribution. }
\label{fig:vertex}
\end{figure}

Given the simple structure of the vertex (\ref{eq:Gamma0}), we can easily sum
the Hartree-Fock diagrams for the single-particle Green's function (cf.\ Fig.\
\ref{fig:vertex}(b)) with the help of the Dyson equation
\begin{align}
\label{eq:Dyson}
\hat{G}^{\Sigma^{(1)}}_{\alpha k,\alpha' k'}(\omega)
&=\hat{G}^{(0)}_{\alpha k,\alpha' k'}(\omega)\\
&\hspace{-25pt}+\sum\limits_{\beta,\beta'}\int\frac{dq}{2\pi}\frac{dq'}{2\pi}
\hat{G}^{(0)}_{\alpha k,\beta q}(\omega)
\hat{\Sigma}^{(1)}_{\beta q,\beta' q'}(\omega)
\hat{G}^{\Sigma^{(1)}}_{\beta' q',\alpha' k'}(\omega).\nonumber
\end{align}
The first-order (Hartree-Fock) contribution to the
self-energy $\hat{\Sigma}^{(1)}$ assumes the following form (we use the
Keldysh matrix notation and suppress the spin-indices as the self-energy is
$\propto \delta_{\sigma'\sigma}$)
\begin{align}
\hat{\Sigma}^{(1)}_{\alpha_1' k_1',\alpha_1 k_1}(\omega)
&=
\frac{1}{\hbar}
\sum_{\alpha_2,\sigma_2}\int\frac{dk_2}{2\pi}
\Gamma_{1' 2,2 1}n_{\alpha_2}(k_2)\hat{\sigma}_{z}\nonumber\\
&=
\frac{\tilde{U}}{\hbar}
a_{\alpha_1'}^\ast \phi_{k_1'}^\ast
\phi_{k_1}^{\phantom{\ast}}a_{\alpha_1}^{\phantom{\ast}} \hat{\sigma}_{z},
\end{align}
with $\tilde{U}=U_{0}N_{\rm int}/2$ and $N_{\rm int}$, cf.\ Eq.\
(\ref{eq:Nint}), is the number of electrons on the level,
\begin{align}
N_{\rs int}=\!\int\!\frac{d(q/\gamma)}{2\pi}
\frac{
(1-\eta)n_{Lq}
+(1+\eta)n_{Rq}}
{1+(\Delta q/\gamma)^2}.
\end{align}
The self-energy is due to the Hartree interaction between electrons of
opposite spins; the Hartree and Fock interaction between equal electrons
cancel each other.

The Dyson equation (\ref{eq:Dyson}) involves the non-interacting Green's
function Eq.\ (\ref{eq:g0}) transformed to $k$-space which is diagonal in the
momentum and lead indices, $\hat{G}^{(0)}_{\alpha k,\beta q}(\omega)
=2\pi\delta(k-q)\delta_{\alpha \beta}\hat{g}^{\rs (0)}_{\alpha k}(\omega)$ with
$\hat{g}^{\rs (0)}_{\alpha k}(\omega)$ given by Eq.~(\ref{eq:gk}).
We solve the Dyson equation (\ref{eq:Dyson}) for $\hat{G}_{\alpha k,\alpha'
k'}^{\Sigma (1)}(\omega)$ in $k$-space and transform the result back to real
space. The real-space Green's function $\hat{G}^{\Sigma (1)}(x,x',\omega)$
then assumes the same form as for the non-interacting case (cf.\
$\hat{G}^{(0)}(x,x',\omega)$ as given by Eq.~(\ref{eq:g0})) but with
renormalized scattering coefficients
\begin{align}
\tilde{t}_{\alpha k}&=t_{\alpha (k-\tilde{U}/\hbar v_F)},\\
\tilde{r}_{\alpha k}&=r_{\alpha (k-\tilde{U}/\hbar v_F)}.
\end{align}
This renormalization of the scattering amplitudes then corresponds to a
mean-field shift of the resonance by $\tilde{U}/\hbar v_{F}$; the general
results Eqs.\ (\ref{eq:scaHartree}) and (\ref{eq:scaFock}) correspond to the
first order expansion in $U_0$ of this result. Higher-order terms in the
self-energy may not preserve the structure of the non-interacting Green's
function, in which case they cannot be cast into a simple renormalization of
the scattering coefficients.

The single-particle Green's function calculated above provides us with the
corrections to the current $j$ and the reducible current-current correlator
$S_{\sigma\sigma}^{\rm (red)}$, cf.\ Eqs. (\ref{eq:jG}) and (\ref{eq:ccri}).
The non-interacting results for these two quantities, Eqs.~(\ref{eq:j0j1}) and
(\ref{eq:cc0}), are modified by substituting the transmission probability
$T_k$ with the expression $\tilde{T}_k=T_{(k-\tilde{U}/\hbar v_F)}$ shifted in
energy due to the interaction. For zero temperature ($\vartheta=0$)
and in the linear-voltage regime, we obtain
\begin{align}
j^{\Sigma^{(1)}}|_{\vartheta=0,V\rightarrow 0}&=\frac{2e^2}{h}\tilde{T}_{k_F} V, \\
S_{\sigma\sigma}^{{\rm (red)}\,\Sigma^{(1)}}|_{\vartheta=0,V\rightarrow 0}&=
\frac{e^2}{h}(1-\tilde{T}_{k_F})\tilde{T}_{k_F} |eV|.
\label{eq:S1ss}
\end{align}

The irreducible current-current correlator $S_{\sigma\pi}^{\rm (irr)}$ is
determined by the two-particle Green's function. Here, we stay with the simple
first-order correction Eq.\ (\ref{eq:cc2p}), as a proper resummation involves
more complex diagrams including vertex corrections.  The irreducible
contribution to the current-current correlator is limited to opposite spins
(where the reducible part does not contribute) and we obtain the result
\begin{align}\label{eq:S1ud}
&S_{\uparrow\downarrow}^{\s (1)}(0)\bigr|_{\vartheta=0}
=e^2 \frac{4U_{0}}{\hbar}\eta(1-\eta^2)^2
\text{sign}(\mu_{\rs L}-\mu_{\rs R})\\
&\hspace{25pt}
\times\int\limits_{k_{\rs min}}^{k_{\rs max}} \!\!
\frac{dk_1 dk_2}{(2\pi)^2}\frac{1}{\gamma^2}
\frac{\Delta k_1/\gamma}
{(1+(\Delta k_1/\gamma)^2)^2(1+(\Delta k_2/\gamma)^2)^2},
\nonumber
\end{align}
which for small voltage reduces to
\begin{align}
S_{\uparrow\downarrow}^{\s (1)}(0)\bigr|_{\vartheta=0,V\rightarrow0}
&=e^2
\frac{4U_{0}}{\hbar}\eta(1-\eta^2)^2 \
\Bigl(\frac{eV}{\hbar v_{\rs F}\gamma}\Bigr)^2
\text{sign}(V)\nonumber\\
&\hspace{40pt}\times\frac{\Delta k_{\rs F}/\gamma} {(1+(\Delta k_{\rs F})^2)^4}.
\end{align}
Its sign 
\begin{align}\label{eq:symm}
\text{sign}(S_{\uparrow\downarrow}^{\s (1)}(0)\bigr|_{\vartheta=0,V\rightarrow 0})
&=\text{sign}(\eta)
\text{sign}(V)\text{sign}(\Delta k_{\rs F})
\end{align}
can be understood on the basis of the mechanism described at the end of Sec.
\ref{sec:irred_corr}.  The result (\ref{eq:S1ss}) provides the lowest order in
$V$ and $U_0$ result for the equal-spin noise correlator; the opposite-spin
correlator is generated by the combination of interaction and asymmetry and is
given by Eq.\ (\ref{eq:S1ud}).

It is instructive to compare our work with the results of Gogolin and 
Komnik~\cite{gogolin:06}. These authors calculated the logarithm of the
generating function $\log \chi$ for the full counting statistics (FCS) in the Anderson impurity model to
second order in the interaction parameter $U/\Gamma$ (with $\Gamma$ the width
of the level), using a particular diagrammatic technique, in linear response
($V\rightarrow 0$) with the following results: i) The FCS assumes the same
form as for non-interacting electrons but with renormalized transmission
probabilities. ii) In the non-linear regime (finite voltage $V$), spin-pairing
correlations are generated to order $V^2$ in bias and to order $(U/\Gamma)^2$
in interaction (pairing correlations have been found \cite{sauret:04} as well
for a dot described within a master equation approach valid at large
interaction $U$).  These results are obtained for a symmetric dot. In our
work, we extend the analysis to an arbitrary scatterer with capacitive-like
Coulomb interaction in the scattering region. Rather than FCS, we limit
ourselves to the calculation of the first two moments (current and noise) of
the FCS and to first order effects in interaction. We confirm the
renormalization of the single-particle scattering matrix in the linear
response regime. For finite voltage $V$, we find that pairing correlations for
electrons with opposite spins are enhanced for an asymmetric scatterer, i.e.,
a scatterer with different reflection coefficients $r_{Lk}\neq r_{Rk}$ for
electrons incident from the left (L) and from the right (R) lead, as they
appear already in the {\it first order} in interaction $U/\Gamma$.

\section{Conclusion}\label{sec:sc}

We have extended the scattering matrix approach to include effects of
interaction; while no restrictions have been imposed on the scatterer, our
formalism applies to the special case of an interaction which is constant over
the scattering region.  Under these conditions, the fully interacting
Hamiltonian can be expressed through the non-interacting (Lippmann-Schwinger)
scattering states and scattering coefficients, hence the scattering aspect of
the problem is dealt with in an exact way. On the part of the interaction, no
knowledge is required about the wave function within the interaction region.
The approach can be extended to the case where scattered electrons interact
with electromagnetic and crystal degrees of freedom, photons, plasmons, and
phonons. 

In our analysis above, we have considered a constant interaction kernel and
one may ask, how well this choice approximates the situation when the kernel
is short-ranged.  For the single resonance level discussed in Sec.\
\ref{sec:srl}, the shape of the interaction kernel is not relevant; the matrix
element Eq.\ (\ref{eq:Asrlm}) derives from the expression Eq.\
(\ref{eq:ovlsca}) for the finite scatterer in the limit of vanishing extent
$d$ and any interaction kernel is always long-ranged.  For a finite dot of
dimension $d$, the shape of the interaction kernel $U(x,x')$ becomes relevant
when its width drops below $d$. Here, we discuss the situation with sharp
resonances and wave functions close to those of an isolated dot. As long as
only a single resonance is involved, the modifications can be captured by a
simple renormalization of the interaction strength $U_0$, with the Hartree and
Fock contributions of similar weight.  The Hartree contribution remains rather
robust when including other occupied resonances, with a uniform
renormalization of the interaction strength over the resonances (for a narrow
kernel, the contributions from the lowest resonances, where the width of the
kernel resolves the structure of the wave functions, is modified).  For the
constant kernel, the inter-resonance Fock terms are strongly suppressed as
compared to the Hartree term, a consequence of the approximate orthogonality
of the wave functions; a narrow kernel then enhances the relative importance
of these (actually small) inter-resonance Fock terms.  Overall, changing the
kernel, we expect small effects on the renormalization of the scattering
coefficients, on the transport current, and on the reducible part of the
zero-frequency noise.  For these quantities, the relevant physics involves
electrons passing through one resonance and interacting with all the electrons
on occupied resonances, which can be roughly included into a renormalization
of $U_0$.  Similarly, a small effect is expected in the irreducible part of
the zero-frequency noise as long as only one resonance is involved in the
transport.  However, at larger voltage, more than one resonance contributes to
the irreducible noise correlator and the interaction between electrons passing
through different resonances becomes important; in this case, the specific
enhancement of the inter-resonance Fock terms should be accounted for, e.g.,
via the introduction of additional `cross-capacitances'. For an open dot with
wide resonances and a short-range kernel, the wave functions in the dot have
to be found explicitly when calculating the overlap matrix elements
(\ref{eq:ov1}), the number of parameters increases, and the results can no
longer be expressed through the scattering coefficients alone in the simple
way we have done above.

As a first application of our approach, we have determined the one- and
two-particle Green's functions within first order perturbation theory in the
interaction parameter $U_0$ for a general scatterer -- an analysis to higher
order is more easily carried out for special scatterers, e.g., a single
resonance level where the interaction Hamiltonian factorizes in $k$-space.
While the linear-response corrections to the one- and the reducible part of
the two-particle Green's functions can be cast into a renormalization of the
scattering coefficients, additional terms of different form show up in the
irreducible part of the two-particle Green's function. In the non-linear
regime at high voltage, corrections appear which are beyond renormalization of
the scattering matrix. Similarly, the interaction-induced corrections to the
current and reducible part of the noise are accounted for through (the same)
renormalized scattering coefficients. In addition, the interaction generates
an irreducible contribution to the noise correlator which goes beyond
renormalization of scattering corefficients. The latter gives rise to
opposite-spin current correlations already to first order in $U_0$ for the
case of an asymmetric scatterer, an effect which has not been addressed so far.

Our results are in agreement with those of Gogolin and Komnik
\cite{gogolin:06} who reported a similar renormalization (in linear response
and to second order in $U_0$) of scattering coefficients in their study of the
full counting statistics of the (symmetric) Anderson model; their spin-pair
correlation appears only in second order of the interaction as their scatterer
is symmetric.  While our study is first order in $U_0$, it applies to any
scatterer with arbitrary scattering coefficients.  We hope, that the
simplifications which come with our approach can be used to generate new (as
with the asymmetric scatterer) or more precise (as with the resummation of a
class of diagrams for the single resonance level) results in other situations.

We thank Andreas Komnik for illuminating discussions and acknowledge financial
support from the Swiss National Foundation, the Pauli Center for
Theoretical Physics, and from the RFBR Grant \# 11-02-00744-a.

\end{document}